\documentclass[11pt,a4paper]{article}
\usepackage{jheppub,amsmath,  amssymb,slashed,url,bm,textgreek,upgreek}
\usepackage{graphicx}
\usepackage{epstopdf}
\usepackage{array}
\usepackage{mathtools}

\usepackage{etoolbox}
\newcommand{\LCDM}{$\Lambda$CDM}
\newcommand{\Le}{Lema{\^{\i}}tre\ }
\newcommand{\ww}{$w_0w_a$}
\title{$\Lambda$CDM:  The path forward}

 \author{Michael S. Turner}
\affiliation{Kavli Institute for Cosmological Physics, The University of Chicago, Chicago, IL  60637-1433}
\affiliation{Department of Physics and Astronomy, University of California, Los Angeles, Los Angeles, CA 90095-1547}
\emailAdd{mturner@uchicago.edu}
\abstract{The current cosmological paradigm, $\Lambda$CDM, is characterized its expansive description of the history of the Universe, its deep connections to particle physics and the large amounts of data that support it.  Nonetheless, $\Lambda$CDM's critics and boosters alike agree on one thing:  it is the not the final cosmological theory and they are anxious to see it replaced by something better!  After reviewing some of the impactful events in cosmology since the last \Le Workshop, I  focus on the role that the recent evidence for evolving dark energy may play in getting cosmology that better theory.
}

\begin{document}
\maketitle

\section{Introduction} 

\begin{figure}
\center\includegraphics[width=0.75\textwidth]{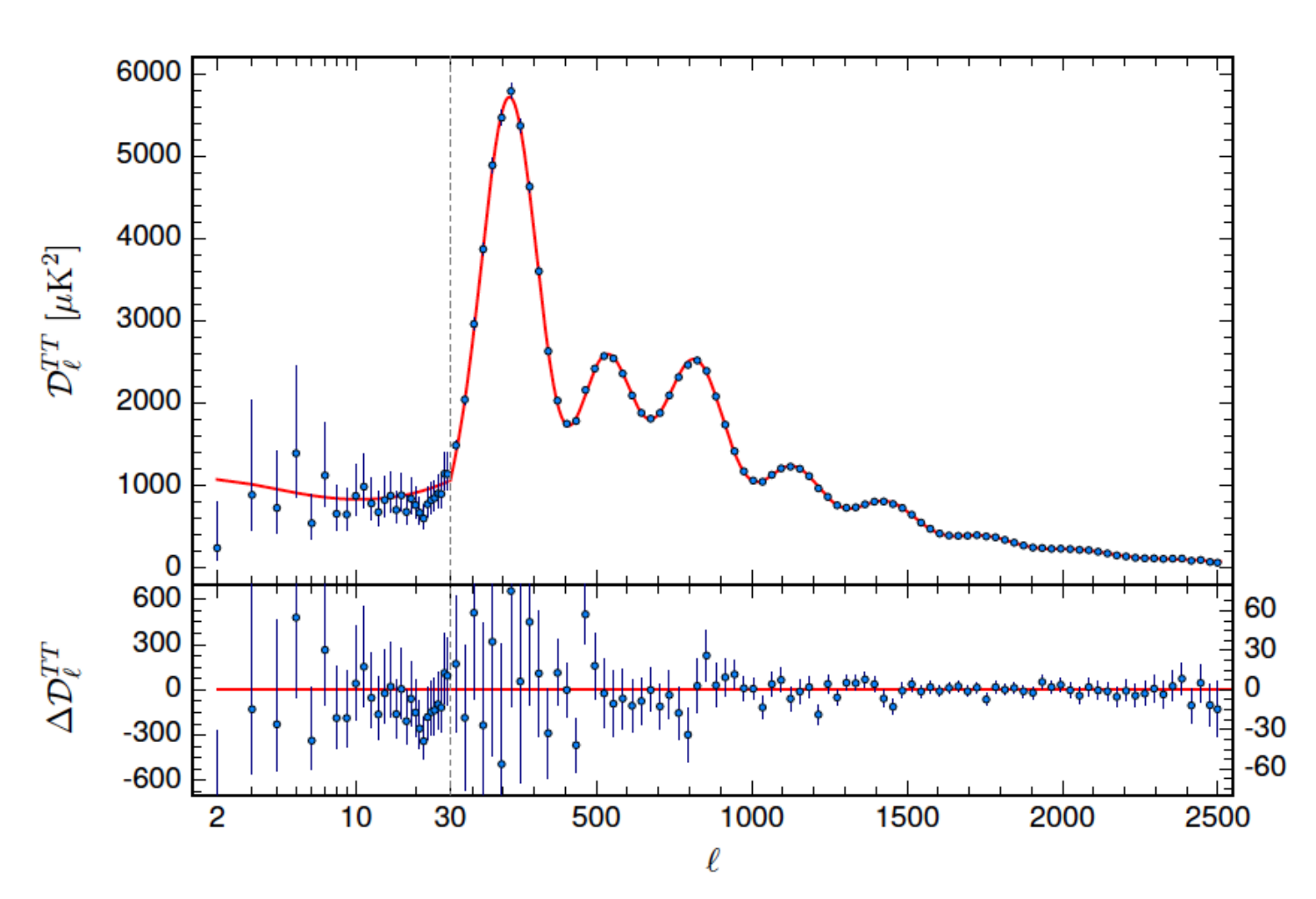}
\caption{CMB temperature anisotropy power spectrum as measured by Planck \cite{PlanckLegacy}; the curve is the best-fit $\rm\Lambda$CDM model.  The concordance of $\Lambda$CDM with the precision CMB measurements is impressive and perhaps the strongest testimony to \LCDM.}
\label{PlanckCMB}
\end{figure}

\LCDM\  is widely known for tracing cosmic history back to a very early inflationary beginning ($t \ll 10^{-6}$\,sec) through the formation of galaxies to today's accelerated expansion, and, for its many successes, most notably agreement with precision measurements of the cosmic microwave background (CMB); see Fig.~\ref{PlanckCMB}.   

The three pillars of $\Lambda$CDM - dark matter, dark energy and inflation -  illustrate the deep connections between the very big, cosmology, and the very small, particle physics.  This connection was a paradigm shift that took place at the end of the twentieth century \cite{Q2C} and remains a hallmark of both cosmology today and particle physics today.  The trio is not only central to $\Lambda$CDM and moving forward, but also they are among the biggest mysteries in both fields today.  

For all its successes, $\Lambda$CDM also has its critics who argue on a regular basis that it has been falsified or must be discarded for various reasons, from the existence of early, massive galaxies revealed by the James Webb Space Telescope to the theoretical shortcomings of inflation \cite{Penrose,PJS}. Critics and boosters alike agree on one thing, though for different reasons:  \LCDM\ needs to be replaced by something better \cite{TurnerPNAS}.  I agree and hope that the successor to $\Lambda$CDM comes even closer to being a ``first principles'' theory than it is now.

$\Lambda$CDM has a first principles foundation, General Relativity and the Standard Model of particle physics, and, the phenomenological add ons, dark matter, dark energy and inflation. By identifying the dark-matter particle and its place in a grander particle physics theory, by providing a fundamental explanation for cosmic acceleration, and by tying inflation to a specific scalar field, $\Lambda$CDM could be ``upgraded'' to a fundamental theory akin to the Standard Model of particle physics.  Or, the needed change to get to the better theory could be more revolutionary, with dark matter and dark energy being eliminated by a new theory of gravity.

After a quick review of important events in cosmology since the last Lema{\^{\i}}tre Workshop, I will focus on the role that the recent evidence for evolving dark energy \cite{DESI} could play in moving cosmology forward.  The Hubble tension could also play that role and others at the meeting will discuss that possibility.

\subsection{Since \Le 2017}
Since the 2017 \Le meeting, much has happened in cosmology relevant to the path forward for $\Lambda$CDM.  The James Webb Space Telescope (JWST) was successfully launched and began science operations in 2022, opening our eyes to  the first stars and galaxies.  Some of the papers featuring early results from JWST suggested that too many massive galaxies existed at high redshifts ($z>10$) to be consistent with $\Lambda$CDM \cite{KillCDM1,KillCDM2}.  However, by now, the misplaced claims of troubles for \LCDM\ have all but disappeared.

In some cases, the redshifts were misidentified; in other cases, galaxy masses were overestimated.  Because $\Lambda$CDM is a theory about the distribution of mass and energy - and not light - and light and mass are not easy to connect, the power of JWST to decisively probe $\Lambda$CDM is limited.

In any case, JWST is revealing much about the first billion years of cosmic history: uninhibited star formation, lots of small, messy galaxies emitting lots of UV light to re-ionize the Universe, and plenty of galaxies with redshifts $z>10$, but no evidence for galaxies with redshifts $z > 20$.

In spite of great effort and sufficient sensitivity to find evidence for a WIMP-like dark matter particle in deep, underground dark-matter searches, at the LHC, and through dark-matter halo-annihilation products, all the searches have come up short.  While there is airtight evidence for nonbaryonic dark matter,\footnote{Namely, the greater than $100\sigma$(!) difference between the total mass density and the baryon mass density as determined by CMB and BBN data.  In fact, the statistical significance is so great that the necessity of non-baryonic dark matter is a fundamental test of \LCDM.} the dark matter particle(s) remains elusive.

It does appear that the neutrino contribution has been nailed down:  The DESI \cite{DR2} upper limit to the sum of the neutrino masses runs up against the lower limit to the sum of neutrino masses from neutrino-oscillation experiments, implying that neutrinos contribute between 0.1\% and 0.15\% of critical density, not so different from the 0.5\% contributed by stars.

The Keck/BICEP CMB experiment at the South Pole continues to be the most sensitive search for inflation-produced gravitational waves, through their B-mode polarization signature \cite{KeckBICEP}. Thus far, it has only set an (impressive) 95\% limit of $r<0.036$.  The Keck/BICEP Collaboration will keep searching; in a year or two from now, they hope to improve their reach by another factor of two.   

On a longer timescale, the CMB-S4 program in the U.S.\footnote{Since the 2024 workshop, DOE and NSF announced that they are no longer supporting the joint Chile/South Pole project, but will however continue to support CMB experiments to achieve the science goals of CMB-S4, including the search for inflation-produced B-mode polarization.} has the goal of achieving a sensitivity of $r \simeq 10^{-3}$,  and a Japanese satellite mission, LiteBird, has a similar aspiration.

The two big developments since the last \Le Workshop that bear on \LCDM\  are the continued Hubble tension, which others at this meeting have discussed, and the DESI results, which provide evidence that dark energy is evolving \cite{DESI,DR2}.  Either (or both) could ultimately be the path to going beyond \LCDM.

I will focus on the later, which may provide a clue about dark energy -- and vindicate my coining the term dark energy in 1998 in order to keep the door open to something other than $\Lambda$ as the explanation for cosmic acceleration \cite{TurnerStromlo}.  What I have to say about the DESI results is largely based upon collaborative work with Abreu \cite{MAMST}.

\section{Evolving dark energy?}
Dark energy can be characterized by an equation-of-state (EOS) parameter $w$ \cite{TurnerWhite}, that may or may not evolve with time, and the evolution of its energy density is given by,
\begin{equation}
d \ln \rho_{DE} = -3(1+w) d \ln a ,
\end{equation}
where $a$ is the cosmic scale factor.  In the case that $w$ is constant, $\rho_{DE} \propto a^{-3(1+w)}$.  In \LCDM, the dark energy is quantum vacuum energy ($\Lambda$) with an unvarying $w=-1$, so that $\rho_{DE} = $ const.
The data -- including the recent DESI results \cite{DESI,DR2} -- are consistent with $w = -1$, with an uncertainty of between $\pm 0.03$ and $\pm 0.1$ \cite{PlanckLegacy,Pantheon,DES5}.   However, there is no compelling theoretical reason to prefer this hypothesis, and its smallness is one of the great mysteries of particle physics and cosmology \cite{WeinbergRMP}.  

The hypothesis of quantum vacuum energy can be falsified by showing that $w \not= -1$, that $w$ evolves with time, or both.  

\begin{figure} [h]
\center\includegraphics[width = 0.75\textwidth]{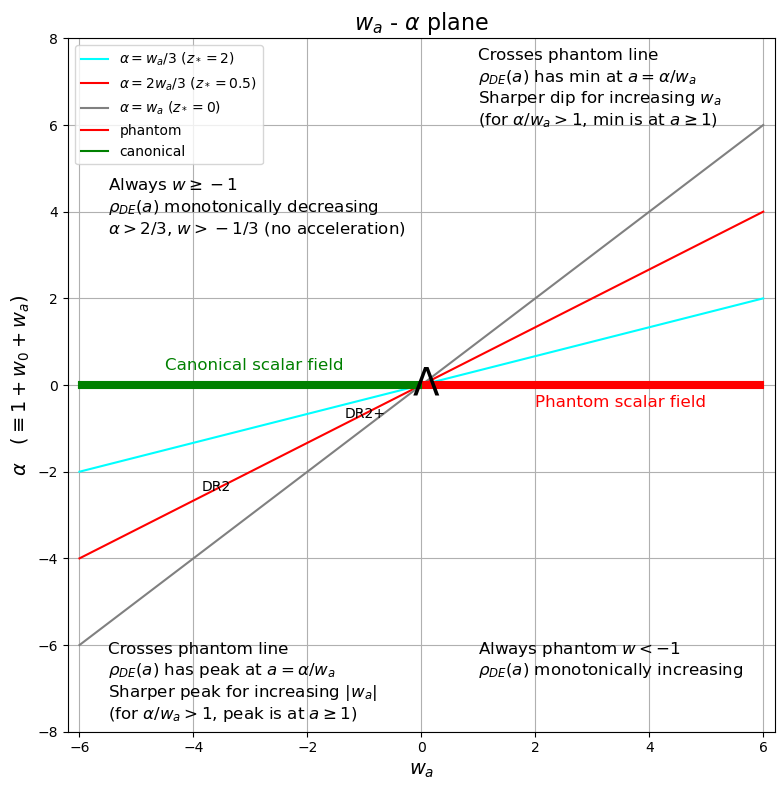}
\caption{The four quadrants of the $\alpha$-$w_a$ plane that define the qualitative behavior of dark energy in the $w_0w_a$ parameterization.  $\Lambda$ corresponds to $\alpha = w_a = 0$, a canonical scalar field is described by $\alpha =0$ and $w_a <0$ and a phantom scalar field by $\alpha =0$ and $w_a > 0$.  DR2 and DR2+ refer to the best fit models to DESI DR2 with (DR2+) and without (DR2) the SNe and CMB data; both are in the SW quadrant where $\rho_{DE}$ has a peak (at $z\simeq 0.5$) and $w$ crosses the phantom line.}
\label{quads}     
\end{figure}

A standard parameterization for $w$ is ``\ww." It is frequently used to probe dark energy by searching for deviations from $\Lambda$ \cite{CP,EL}.  Since the DESI evidence for evolving dark energy is described in terms of $w_0w_a$, I will describe it in a little more detail.  Briefly, 
\begin{eqnarray}
 w & = & w_0 + w_a(1-a) = w_0 +w_a z/(1+z) \nonumber \\
 & = & -1 +\alpha -w_a a ,
 \end{eqnarray}
where $\alpha \equiv 1+w_0+w_a$, $a = 1$ today, and the cosmic scale factor and cosmological redshift are related by: $a(t) = 1/(1+z)$.   
In this parameterization:
\begin{itemize}
\item The value of $w$ today is $w_0$
\item At early times ($a\ll 1$) $w\rightarrow  w_0+w_a$
\item At late times ($a\gg 1$) $w \rightarrow -a w_a$
\end{itemize}
Further, it follows from Eq.~(1) that the energy density of dark energy is a power-law times an exponential, whose behavior is determined by $w_a$ and $\alpha$:
\begin{equation}
\rho_{DE} \propto a^{-3\alpha} \exp [3w_a (a-1)].
\end{equation}
The generic behavior of $\rho_{DE}$ depends upon which ``quadrant'' $\alpha$ and $w_a$  occupy:
\begin{itemize}
\item{NE ($\alpha , w_a >0$):} $\rho_{DE}$ achieves a minimum for $a = \alpha / w_a$ and crosses the phantom line ($w=-1$)
\item{SW ($\alpha , w_a <0$):} $\rho_{DE}$ achieves a maximum for $a = \alpha / w_a$ and crosses the phantom line ($w=-1$)
\item{NW ($\alpha >0$, $w_a < 0$):} $\rho_{DE}(a) $ monotonically decreases and $w \ge -1$ always
\item{SE ($\alpha <0$, $w_a> 0$):} $\rho_{DE}(a) $ monotonically increases and $w \le -1$ always
\end{itemize}
$\Lambda$ corresponds to the singular case: $w_0 =-1$, $w_a = \alpha = 0$, where $\rho_{DE} \propto \  const$.  Further, if dark energy is a rolling scalar field with a canonical kinetic term, $w \rightarrow -1$ as $a\rightarrow 0$, and so $\alpha$ must be zero and $w_a < 0$.  For a phantom scalar field (negative kinetic term), $w \rightarrow -1$ as $a\rightarrow 0$, so $\alpha$ must be zero and $w_a >0$.  The $\alpha$-$w_a$ plane is summarized in Fig.~\ref{quads}.  

While the \ww\ parameterization has proven very useful, it is agnostic about physics, and $\rho_{DE}$ can have unphysical behavior.  For example, $w<-1$ (phantom dark energy) for $w_a > 0$ and $\alpha <0$; $w$ crosses the $w=-1$ phantom divide for $w_a >0$ and $\alpha >0$ or $w_a <0$ and $\alpha <0$ (the case favored by DESI).

\subsection{DESI, CMB and SNe results}
By determining the redshifts of more than 20 million galaxies, the DESI Collaboration has measured percent-level BAO distances out to redshift $z = 4$ \cite{DESI,DR2}.  Using the $w_0w_a$ parameterization and combining their results with CMB and SNe data \cite{PlanckLegacy,DES5,Pantheon}, they found $3\sigma$ to $4\sigma$ evidence for varying dark energy.  In particular, $w_0 \simeq -0.7$ and $w_a \simeq -1$ ($\alpha \simeq -0.7$) is a statistically-significant, better-fit to the DESI + CMB + SNe data than $\Lambda$CDM  \cite{DESI,DR2}.

\begin{figure}[h]
\center\includegraphics[width = 0.7\textwidth]{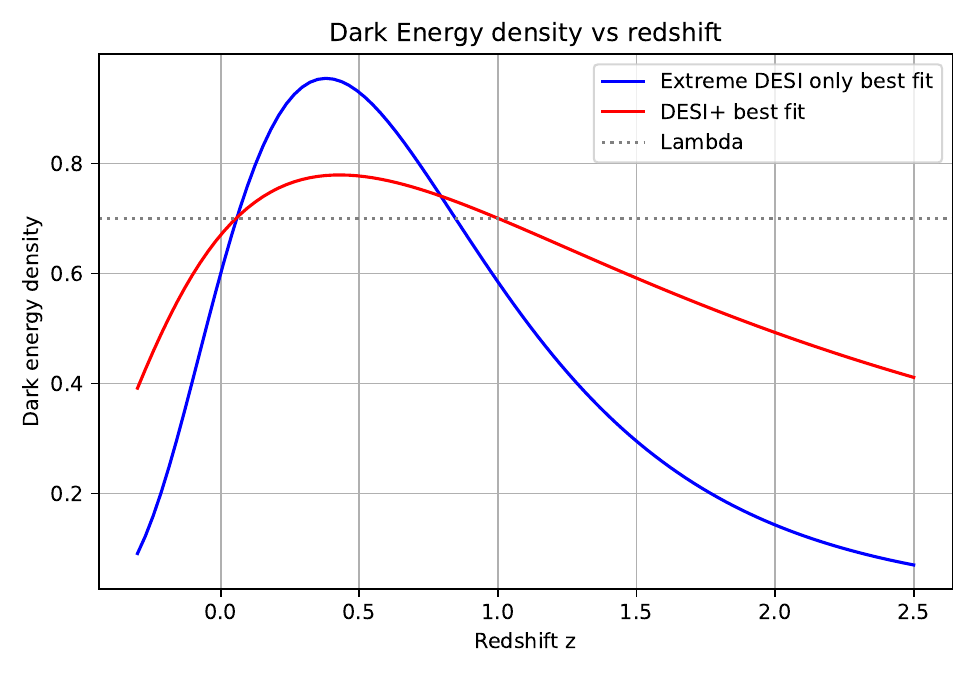}
\caption{Evolution of the energy density of dark energy (in units of the critical density today) for the extreme DESI-data only best-fit (blue), the DESI+ best-fit (red), and $\Lambda$.  The extreme DESI model has $\Omega_M = 0.4$, $w_0=0.016$, and $w_a = -3.69$.  The DESI+ best-fit includes SNe and CMB data, and has $\Omega_M = 0.33$, $w_0=-0.7$, and $w_a = -1$ }
\label{rhoDE}     
\end{figure}

For this model, $w$ increases from $-1.7$ at high redshift, crosses the phantom line around $z\simeq 1/2$, and asymptotically approaches large positive values in the future.  Further, $\rho_{DE}$  achieves a maximum around $a = \alpha /w_a \simeq 2/3 $, or $z\simeq 1/2$, and decreases rapidly earlier and later; see Fig.~\ref{rhoDE}.  To say the least, this is surprising behavior!

Considering the DESI BAO data alone, there is a degeneracy direction in the $w_0w_a$ likelihood plane: $w_a \simeq -3(1+w_0)$ or $\alpha \simeq {2\over 3}w_a$ \cite{DESI,DR2}.  This degeneracy direction persists when the CMB and SNe are included as well. That means that the DESI data tightly constrain the position of the peak in $\rho_{DE}$, at $a = \alpha /w_a \simeq 2/3$ or $z \simeq 0.5$, but not the value of $w_a$, which controls the width of the peak. 

\begin{figure} [h]
\center\includegraphics[width = 0.75\textwidth]{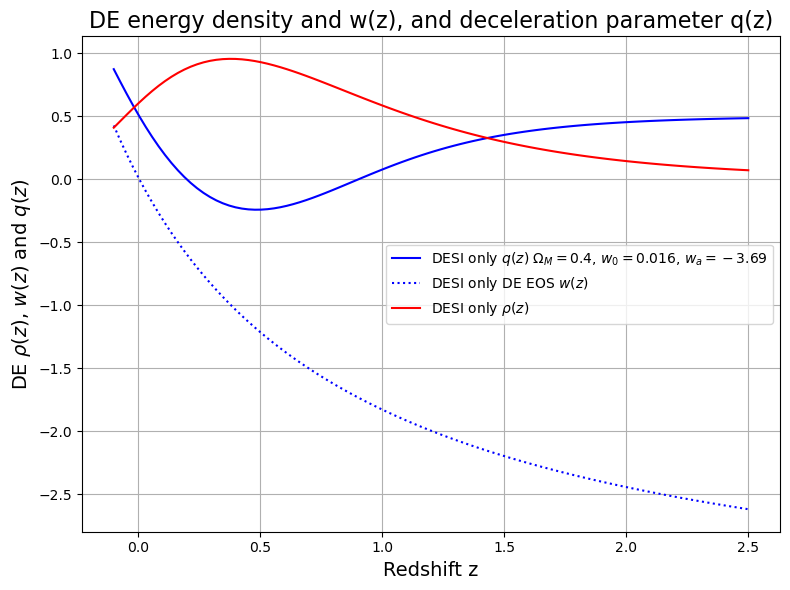}
\caption{Best-fit, DESI-only data $w_0w_a$ model ($\Omega_M = 0.4$, $w_0 = 0.016$ and $w_a = -3.69$):  Energy density of dark energy (red), dark energy EOS (dotted blue) and deceleration parameter (blue).  Note, in this model the Universe is not accelerating today, $q_0 \simeq 0.5$.}
\label{EOSq}     
\end{figure}

Figure \ref{EOSq} shows the very-peaked energy density of dark energy and its equation-of-state for the extreme DESI model.  The EOS crosses the phantom divide around $z \simeq 0.5$ and the Universe is no longer accelerating today and $q_0 \simeq 0.5$!

If the DESI+ best fit $w_0w_a$ model (or the extreme DESI model) reflects reality, we live at a very special time, shortly after the dark energy suddenly appears and then disappears.  While $w_0w_a$ has become a standard parameterization for probing dark energy, it has the shortcomings mentioned above.  For this reason, Abreu and I  explored physics-based models -- rolling scalar fields \cite{MAMST} -- which I will now discuss.  

\subsection{Scalar-field dark energy}
A scalar field which is displaced from the minimum of its potential and is initially stuck due to Hubble friction is a model for evolving dark energy \cite{FHSW,PR}.  When the scalar field begins to roll, $w$ increases from $w= - 1$, at a rate determined by the shape of the potential.  The coupled equations for the evolution of the cosmic scale factor $a(t)$ and a scalar field $\phi$ with canonical kinetic energy are:
\begin{eqnarray}
H^2 \equiv (\dot a/a)^2 & = & {8\pi (\rho_M + \rho_\phi  )\over 3 m_{\rm pl}^2} \\
\rho_\phi & = & {1\over 2}\dot\phi^2 + V (\phi ) \\
p_\phi & = & {1\over 2}\dot\phi^2 - V (\phi ) \\
w_\phi &\equiv & p_\phi / \rho_\phi \\
0 & = &\ddot\phi + 3H\dot \phi + \partial V (\phi )/\partial \phi
\end{eqnarray}
where $\hbar = c =1$, $G=1/m_{\rm pl}^2$, and a spatially-flat Universe is assumed. A phantom scalar field has a negative kinetic term, and the scalar field rolls uphill, with $w$ decreasing from $-1$.

We considered three scalar potentials: a massive scalar field, $V(\phi )= m^2\phi^2/2$, a free scalar field, $V(\phi )= \phi^4/4$, and an exponential potential (motivated by the moduli fields of string theory), $V(\phi )= V_0 \exp [\beta \phi /m_{\rm pl} ]$, as well as a massive phantom scalar field.  We showed that the quartic and exponential potentials can be made equivalent to a massive scalar field for the parameters relevant to the DESI results, and so I will only discuss results for a massive scalar field.

A useful dimensionless parameter for the massive scalar-field model is:
$$\beta \equiv m^2/H_0^2,$$
where $H_0$ is the present value of the Hubble parameter.  In the limit that $\beta \rightarrow 0$, the massive scalar field model becomes \LCDM.  Increasing $\beta$ corresponds to earlier rolling and more deviation from $\Lambda$.  (For the quartic and exponential potentials, there is a similarly-defined dimensionless quantity that plays the same role as $\beta$.)

Once the potential is specified, scalar-field models have one more parameter than $\Lambda$CDM, namely $\beta$, which is one less parameter than a $w_0w_a$ model.  To compare the models to the first DESI data release (DR1) \cite{DESI}, we use the $\chi^2$ statistic for the 12 distances that DESI measured.  For \LCDM, $\chi^2 = 12.8$, and for the DESI model preferred by DESI + CMB + SNe, $\chi^2 = 11.3$, which is significantly better.

For each value of $\beta$, we selected the values of $\Omega_M$ and $H_0 r_d$ that minimize $\chi^2$.\footnote{The DESI distances are given in units of the BAO damping scale $r_d$ and in our models the distances are in units of $H_0^{-1}$.  To connect the two, another parameter must be specified:  $H_0r_d$.}   $\chi^2$ increases with increasing $\beta$ and rises above $20$ for $\beta$ greater than a few.  For the non-tachyonic models, the rise is monotonic.  The tachyonic model is more interesting.  The minimum value of $\chi^2$ occurs for $\beta \simeq 0.5$, at value slightly, but not significantly, lower than $\Lambda$CDM, $\chi^2 = 12.5$.

The bottom line is scalar-field dark-energy does not significantly improve the fit to the first-year DESI data as compared to $\Lambda$CDM, and it cannot match the  improvement of the best $w_0w_a$ models.

The second data release of DESI (DR2) \cite{DR2} has 33 distances in total, and in general the results are similar to those for DR1.  $\Lambda$CDM is a good fit with a $\chi^2 = 29.6$.  Once again, a $w_0w_a$ model is a significantly better fit:  with the CMB constraint,\footnote{The CMB tightly constrains $H_0r_d$ through its very precise determination of the sound horizon.} $\Delta\chi^2 = -5.6$ for $\Omega_M = 0.353$, $w_0 = -0.435$ and $w_a = -1.75$.  (Without the CMB constraint, $\Delta\chi^2 = -10$ for $\Omega_M = 0.4$, $w_0 = 0.016$ and $w_a = -3.7$.)  The DESI distance results continue to favor a sharply peaked dark-energy energy density around $z\simeq 0.5$, fixed by the degeneracy direction in the $w_0w_a$ plane, $w_a \simeq -3(1+w_0)$;  cf.~Fig.~11 in Ref.~\cite{DR2}.  

\begin{figure}[h]
\center\includegraphics[width = 0.75\textwidth]{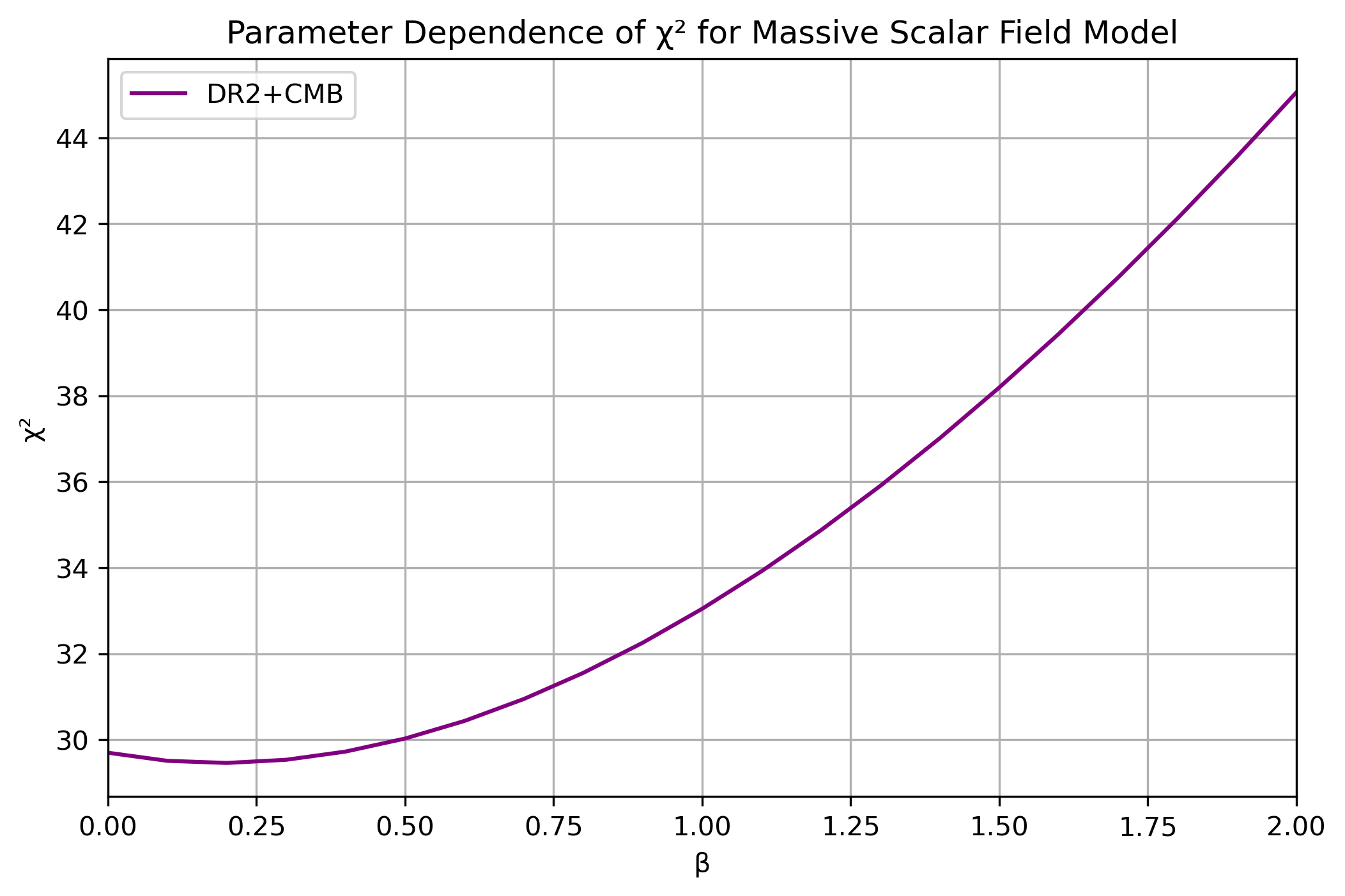}
\caption{$\chi^2$ vs. $\beta$ for the massive scalar field model and DR2 with the CMB constraint and the value of $\Omega_M$ that minimizes $\chi^2$.}
\label{DR2Massive}     
\end{figure}

There are differences between DR1 and DR2.  $\chi^2$ for the massive scalar field model is now minimized for $\beta \simeq 0.2$ (see Fig.~\ref{DR2Massive}).  And secondly, the minimum $\chi^2$ for the tachyonic model is now at $\beta =0$ and $\chi^2$ monotonically -- and steeply -- rises for increasing $\beta$.

The DESI preference for a sharply-peaked dark energy is a robust feature of both DR1 and DR2.  Because of this, we tried two {\it ad hoc} models for $\rho_{DE}$:  a Gaussian of adjustable width $\sigma$ centered on $z = 0.4$ and a log-normal of adjustable width also centered on $z = 0.4$.  The Gaussian is too sharp:  $\chi^2$ is minimized for $\sigma \rightarrow \infty$, which corresponds to $\Lambda$.  While the log-normal model is almost indistinguishable from the best-fit $w_0w_a$ model for $\rho_{DE}$, $\chi^2$ is minimized for $\sigma = 0.35$ with $\Delta \chi^2 = - 8.4$, which is not as good as the best $w_0w_a$ model ($\Delta \chi^2 = -10$).  Whether or not $w_0w_a$ is well-motivated, the DESI data love it!

\subsection{Adding SNe}
Type Ia SNe probe cosmological distances over a similar redshift range as the DESI results do with much less precision per measurement (typically 20\% or so), but with many more measurements, and, most importantly different systematic errors.  We used the Pantheon+ SH0ES data set of 1701 SNe \cite{Pantheon}, which cover redshifts $z = 0.001$ to $z = 2.26$, to constrain our models for evolving dark energy. 

\begin{figure}[h]
\center\includegraphics[width = 0.75\textwidth]{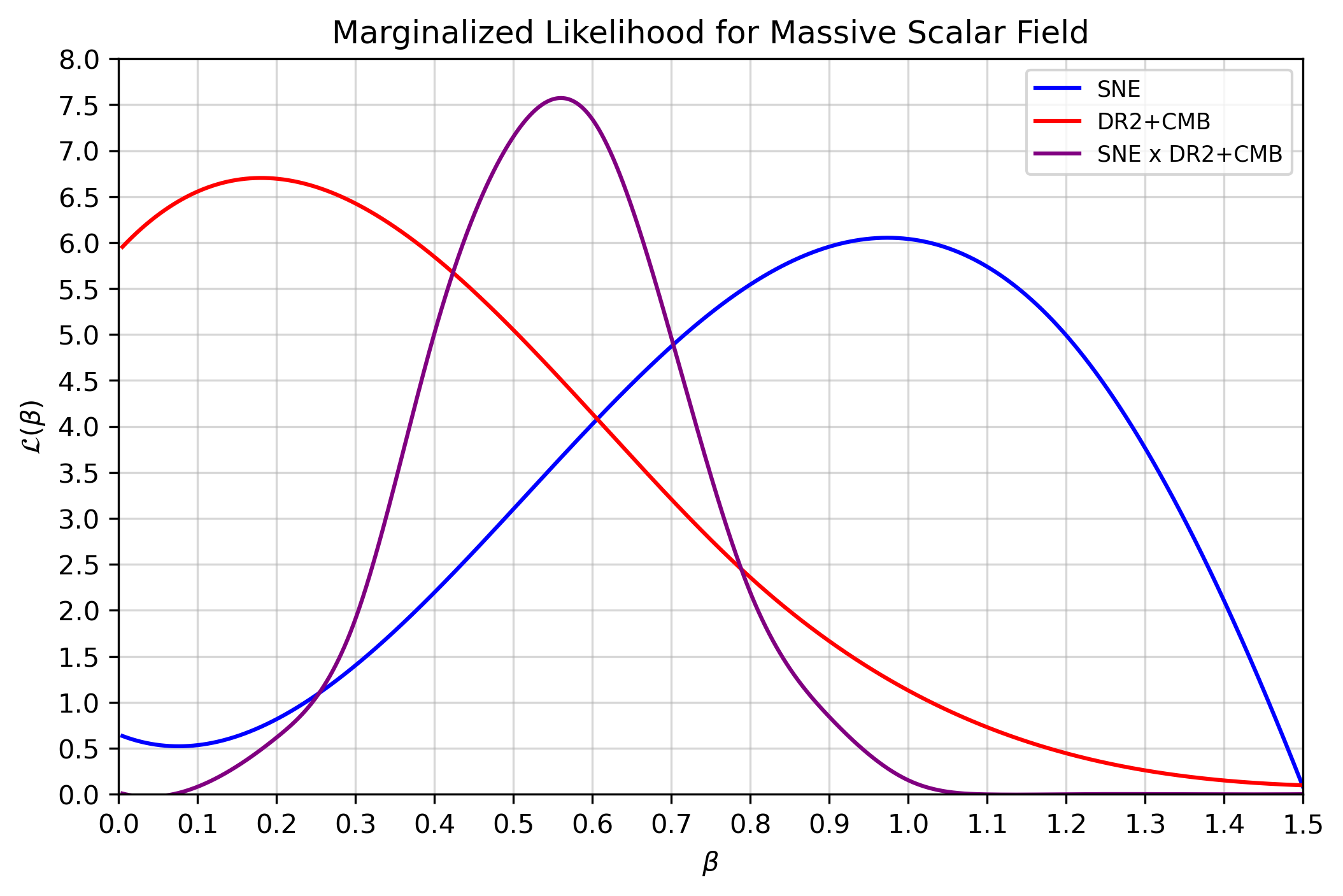}
\caption{Marginalized (over $\Omega_M$) likelihoods for DR2, SNe and the DR2 + SNe dataset.  The 95\% credible ranges are respectively:  $\beta = 0 -1$, $\beta = 0.2 - 1.5$ and $\beta = 0.22 - 0.95$.}
\label{Likelihoods}     
\end{figure}

In comparing our theoretical models to the data, there are three parameters:  $H_0$, $\Omega_M$ and $\beta$.  We have computed the $\chi^2$ values for our models using the Pantheon+ SH0ES covariance matrices, cf. sections 2.2 and 2.3 of Ref.~\cite{Pantheon}.  We have minimized $\chi^2$ by the choice of $H_0$, which is largely determined by 40 some low-redshift supernovae with Cepheid distances; this leaves $\chi^2$ as a function of $\Omega_M$ and $\beta$.\footnote{For our purposes, $H_0$ is a nuisance parameter since we can change the absolute distances and thereby $H_0$ without affecting the dark energy analysis.  We have verified this fact by deleting the 60 or so low-redshift SNe that have Cepheid distances, which does not affect our results.}  From this we have computed the likelihood function, $\mathcal{L} \propto \exp ({-\chi^2}/2)$.  We have also computed the likelihood function for the DR2 dataset as a function of $\Omega_M$ and $\beta$.  The results, marginalized over $\Omega_M$,  are shown in Fig.~\ref{Likelihoods}.  

The SNe results favor a value of $\beta \sim 1$, with a 95\% credible range, $\beta = 0.2 - 1.5$, where $\Delta\chi^2$ decreases to a minimum of around $-7$.  For comparison, the best-ft $w_0w_a$ model is not as good a fit: $\Delta\chi^2 = - 5.8$ for $\Omega_M = 0.4$, $w_0 = -0.72$ and $w_a = - 2.77$.  On the other hand, the DR2 results weakly favor a lower value, $\beta \sim 0.2$, with the 95\% credible range $\beta = 0 - 1.1$.  Also shown in Fig.~\ref{Likelihoods} is the joint DR2/SNe likelihood marginalized over $\Omega_M$.  The 95\% credible range is $\beta = 0.22 - 0.95$.

In sum, DR2 (and DR1) strongly prefer sharply peaked dark-energy (around $z\simeq 0.5$), while the SNe data have a mild preference for a rolling scalar field.  Together, SNe and DR2 provide evidence for scalar-field dark energy and even stronger evidence for a $w_0w_a$ model of evolving dark energy \cite{DESI,DR2}.

\subsection{Take aways}
Bringing it all together:
\begin{itemize}

\item The best-fit $w_0w_a$ DESI+ model ($w_0=-0.7$ and $w_a=-1$) is a slightly better fit to DR1 than $\Lambda$CDM, $\Delta \chi^2 \simeq -1.5$.  It does so at the expense of the energy density of dark energy achieving a maximum around $z\simeq 0.5$, cf. Fig.~\ref{rhoDE}.

\item The DESI results -- both DR1 and DR2 -- favor a dark-energy energy density that has a peak around $z \simeq 0.5$, decreasing at higher and lower redshifts, cf. Fig.~\ref{rhoDE}.\footnote{This also leads to a bump in $H(z)$ relative to $\Lambda$CDM of a few percent around $z\simeq 0.5$, which in turns leads to a step-like decrease in distances around $z\simeq 0.5$.}  A peaked energy density doesn't look like a rolling scalar field or anything I have ever seen!

\item None of the scalar-field models fit as well as the best $w_0w_a$ models.  For DR1 the tachyonic model fits slightly better than $\Lambda$CDM.  For DR2, the massive scalar-field model with $\beta \sim 0.2$ is a slightly better fit than $\Lambda$CDM and the tachyonic model is completely disfavored.

\item The Pantheon+SH0ES type Ia supernovae results favor a massive scalar-field model with $\beta \sim 1$ over a $w_0w_a$ model.  Taken together, DR2, CMB and the Pantheon+SH0ES datasets imply a 95\% credible range $\beta = 0.22 - 0.95$, providing some evidence for scalar-field dark energy.

\item While the $w_0w_a$ parameterization is useful for testing for a deviation from $\Lambda$, it cannot describe scalar-field models to the precision required over the redshift range of the data \cite{MAMST}.  Its inability to accurately describe a rolling scalar field traces to the fact that in fitting a scalar field it is a one parameter fit, cf., Fig.~\ref{quads}.


\end{itemize}

As exciting as the hints of evolving dark energy are, $\Lambda$CDM is a good fit to both DR1 and DR2 as well as a host of other cosmological data.  And it is possible, that as additional data are accumulated, the evidence for a deviation from $\Lambda$CDM will disappear.  Or not!  

Last but not least, one strong message from the DESI high-precision BAO measurements is their preference for a sharply-peaked dark energy described by a $w_0w_a$ model.  Such behavior is not characteristic of any familiar mass/energy component. This robust fact is either telling us something important about dark energy or the DESI data themselves.

\section{Concluding remarks}

Harking back to the 2017 \Le meeting, the title of my talk \cite{Turner2018} was, ``{\it \LCDM:  Much more than we expected, but now less than what we want.}''  One cannot overstate how big a leap forward \LCDM\ was.  


In a few years, cosmology went from the hot big-bang model, which began at big-bang nucleosynthesis, with poor knowledge of the basic large-scale features of the Universe and a paucity of data, to \LCDM, which begins at the tiniest fraction of a second after the big bang, explains all the basic features of the Universe, and is supported by a wealth of precision data.  

What a leap!  If \LCDM's successor is a similar leap in understanding and scope, it is hard to even imagine what it might look like.  Will it address the origin of space, time and the Universe, as well as the destiny?  Will dark matter and dark energy even be the right questions?  And will it clarify or make moot the notion of a multiverse?

The path forward could be evolutionary - discovering the dark matter particle, understanding dark energy at a more fundamental level, and finding a standard model of inflation - or it could be revolutionary - for example, replacing dark matter and dark energy with a new theory of gravity.  The triggering event could be a surprising observation, a discrepancy that won't go away, or a theoretical breakthrough.  At this meeting we have heard about two possible ``triggering events'':  the Hubble tension and the DESI results.  

Linde, one of the originators of inflation and a participant at this meeting, more than once said, you can only kill inflation with a better theory.  While I found his statement jarring the first time I heard it, I think it has some relevance here.  

Given $\Lambda$CDM's many successes and the absence of a competitor that can match even a few of them, it is too early to discard it.  And, because the body of data that supports it is so extensive, its successor will first have to make itself look like $\Lambda$CDM before it reveals its exciting new features.

\vskip0.5cm
 \noindent {\bf {Acknowledgements}}  \\
 I thank organizers of Lema{\^{\i}}tre Conference 2024, where this work was presented. I also thank Dragan Huterer and Eric Linder for enlightening conversations.  Most of the discussion of the DESI results comes from collaborative work with Matilde Abreu, described in Ref.~\cite{MAMST}.


\begin{thebibliography}{99}


\bibitem{Q2C}  {\it Connecting Quarks with the Cosmos: Eleven Science Questions for the New Century} (The National Academies Press, Washington, DC, 2003).
\bibitem{PlanckLegacy} Planck Collaboration, {\it Astron. Astrophys.} {\bf 641}, A1 (2020).
\bibitem{Penrose} R. Penrose, {\it Found. Phys.} {\bf 48}, 1177 (2018)
\bibitem{PJS} P.J. Steinhardt, {\it Sci. Am.} {\bf 304}, 18 (2011).
\bibitem{TurnerPNAS} M.S. Turner, {\it PNAS}, to be published; arXiv: 2510.05483v1 (2025).
\bibitem{DESI} A.G. Adame et al (DESI Collaboration), arXiv:2404.03002v1 (2024).
\bibitem{KillCDM1} I. Labbe et al, {\it Nature} {\bf 616}, 266 (2023)
\bibitem{KillCDM2} M. Boylan-Kolchin, {\it Nature Astronomy}, {\bf 7}, 731 (2023)
\bibitem{DR2}  M.A. Karim et al (DESI Collaboration), arXiv:2503.14738 (2025).
\bibitem{KeckBICEP} P.A.R. Ade et al, {\it Phys. Rev. Lett.} {\bf 127}, 151301 (2021).
\bibitem{TurnerStromlo}  M.S. Turner, in {\it Proceedings of The Third Stromlo Symposium: The Galactic Halo}, eds. B.K. Gibson, T.S. Axelrod, and M.E. Putnam, Astron. Soc. Pac. Conf. Series 165, 431 (1999) (astro-ph/9811454). 
\bibitem{MAMST} M. Abreu and M.S. Turner, arXiv:2502.08876 (2025).
\bibitem{TurnerWhite} M.S. Turner and M. White, {\it Phys. Rev. D} {\bf 56}, R4439 (1997).
\bibitem{Pantheon} D. Brout et al, {\it Astrophys. J.} {\bf 938}: 100 (2022).
\bibitem{DES5} T.M.C. Abbott et al (DES Collaboration), arXiv:2401.02929v2 (2024).
\bibitem{WeinbergRMP} S. Weinberg, {\it Rev. Mod. Phys.} {\bf 61}, 1 (1989).
\bibitem{CP} M. Chevallier and D. Polarski, {\it Int. J. Mod. Phys. D} {\bf 10}, 213 (2001).
\bibitem{EL}  E.V. Linder, {\it Phys. Rev. Lett.} {\bf 90}, 091301 (2003).
\bibitem{FHSW} J. Frieman, C. Hill, A. Stebbins, and I. Waga, {\it Phys. Rev. Lett.} {\bf 75}, 2077 (1995).
\bibitem{PR} P.J.E. Peebles and B. Ratra, {\it Phys. Rev. D} {\bf 37}, 3406 (1988).
\bibitem{Turner2018} M.S. Turner, {\it Found. Phys.} {\bf 48}, 1261 (2018).







\end{thebibliography}
\end{document}